\documentclass[%
reprint,
 amsmath,amssymb,
 aps,
pra,
]{revtex4-2}
\usepackage{physics}
\usepackage{tikz}
\usetikzlibrary{decorations.markings, calc, arrows.meta, positioning, shadings}

\tikzset{laser/.style={thick, black}}
\tikzset{arrow inside/.style = {postaction=decorate, decoration={markings, mark=at position .62 with \arrow{stealth}}}}
\tikzset{arrow inside1/.style = {postaction=decorate, decoration={markings,
						 mark=at position .62 with \arrow{stealth}}}}
\tikzset{ray/.style={very thick, red, arrow inside}}
\tikzset{ray1/.style={very thick, red, arrow inside1}}
\tikzset{detector/.style={thick, draw=black, fill=black!40}}
\tikzset{reflector/.style={thick, black, left color=black!50, right color=black!50, middle color=white}}
\tikzset{reflector1/.style={thick, black, left color=black!50, right color=black!50, middle color=white}}

\usepackage{graphicx}
\usepackage{dcolumn}
\usepackage{bm}
\usepackage[colorlinks=true,linkcolor=blue,citecolor=blue,urlcolor=blue]{hyperref}

\usepackage{caption}
\usepackage{subcaption}

\usepackage{ragged2e}
\usepackage{hyperref}
\DeclareCaptionJustification{justified}{\justifying}

\begin{document}

\preprint{APS/123-QED}

\title{Existence of unbiased estimators in discrete quantum systems}
\author{Javier Navarro}
\email{jnavarro@bcamath.org}
\affiliation{BCAM - Basque Center for Applied Mathematics, Mazarredo 14, 48009 Bilbao, Spain}
\affiliation{Department of Physical Chemistry, University of the Basque Country UPV/EHU, Apartado 644, 48080 Bilbao, Spain}

\author{Ricard Ravell Rodr\'iguez}
\email{ricard.ra.ro@gmail.com}
\affiliation{BCAM - Basque Center for Applied Mathematics, Mazarredo 14, 48009 Bilbao, Spain}
\affiliation{Institute for Cross-Disciplinary Physics and Complex Systems IFISC (UIB-CSIC),
Campus Universitat Illes Balears, E-07122 Palma de Mallorca, Spain}
\author{Mikel Sanz}
\email{mikel.sanz@ehu.eus}

\affiliation{BCAM - Basque Center for Applied Mathematics, Mazarredo 14, 48009 Bilbao, Spain}
\affiliation{Department of Physical Chemistry, University of the Basque Country UPV/EHU, Apartado 644, 48080 Bilbao, Spain}
\affiliation{EHU Quantum Center, University of the Basque Country UPV/EHU, Apartado 644, 48080 Bilbao, Spain}
\affiliation{IKERBASQUE, Basque Foundation for Science, Plaza Euskadi 5, 48009, Bilbao, Spain}
\begin{abstract}
The Cramér-Rao bound serves as a crucial lower limit for the mean squared error of an estimator in frequentist parameter estimation. Paradoxically, it requires highly accurate prior knowledge of the estimated parameter for constructing the optimal unbiased estimator. In contrast, Bhattacharyya bounds offer a more robust estimation framework with respect to prior accuracy by introducing additional constraints on the estimator. In this work, we examine divergences that arise in the computation of these bounds and establish the conditions under which they remain valid. Notably, we show that when the number of constraints exceeds the number of measurement outcomes, an estimator with finite variance typically does not exist. Furthermore, we systematically investigate the properties of these bounds using paradigmatic examples, comparing them to the Cramér-Rao and Bayesian approaches.
\end{abstract}

\maketitle
\section{Introduction}\label{Sec1}
Quantum metrology investigates the fundamental limits of precision in estimating physical parameters encoded in quantum systems. Taking advantage of inherently quantum features such as entanglement and squeezing, it enables precision levels beyond those attainable by classical estimation strategies~\cite{paris,toth2014quantum,giovannetti2011advances}. Applications span quantum phase estimation~\cite{ligo2011gravitational,ligo2013enhanced,demkowicz2015quantum,gessner2018sensitivity}, quantum-enhanced position and velocity estimation~\cite{zhuang2022ultimate,reichert1,reichert3}, quantum illumination~\cite{tan2008quantum,sanz2017quantum,reichert2}, quantum thermometry~\cite{mehboudi2019thermometry,rodriguez2024strongly,mehboudi2019using,binder2018thermodynamics}, and quantum channel discrimination~\cite{davidovich2023quantum}, driving substantial progress in the field. The theoretical framework of parameter estimation provides benchmarks for the optimal performance of quantum-enhanced estimation protocols.

Parameter estimation has traditionally been approached from two main perspectives: frequentist and Bayesian~\cite{casella2002statistical}. These frameworks, based on different interpretations of probability, yield distinct insights into parameter estimates and their uncertainties. Although frequentist and Bayesian approaches converge in the asymptotic limit of large measurement numbers, this has often led to the misconception that they are interchangeable in quantum metrology. In reality, they reflect fundamentally different paradigms. In the frequentist framework, the Cramér-Rao bound (CRB)~\cite{Cramer:107581,Rao1992} is the main figure of merit, providing a lower bound on the mean squared error of estimators. However, achieving the precision dictated by the CRB paradoxically requires either a vast number of measurements or highly accurate prior knowledge of the parameter—both of which are often impractical or experimentally infeasible. In finite-sample regimes, the CRB frequently becomes unattainable~\cite{rubio2018non,meyer2023quantum,jarzyna2015true}. To address these limitations, alternative bounds have been developed, including the Hammersley–Chapman–Robbins bound~\cite{chapman1951minimum}, the Bhattacharyya bounds (BhBB)~\cite{bha}, the Barankin bounds~\cite{barankinthres,knockaert1997barankin}, and the more general Abel bounds~\cite{abel1993bound}. These bounds are particularly valuable in noisy estimation scenarios, providing more realistic precision limits~\cite{knockaert1997barankin}.

Recent efforts have aimed to derive tighter bounds for quantum parameter estimation.  The first result on this line was published by Tsuda~\cite{tsuda2007bhattacharyya}.
\color{black} More recently, Gessner and Smerzi~\cite{Gessner} proposed a hierarchy of lower bounds on estimator variance by imposing additional unbiasedness constraints. This hierarchy includes the quantum CRB (QCRB)~\cite{braunstein1994statistical} as its lowest-order bound, while higher-order cases include quantum analogs of the Barankin, Bhattacharyya, and Abel bounds.

In this work, we analyze both classical and quantum Bhattacharyya bounds (QBhBB) and assess their performance relative to the CRB and Bayesian approaches, particularly in scenarios with uncertain prior knowledge of the parameter. We first illustrate these bounds with two paradigmatic cases, namely, the estimation of the variance of a Gaussian distribution, and the analysis of the performance of a Mach-Zehnder interferometer setup. Our findings reveal that the BhB-inspired strategy can provide a significant advantage in specific regimes. Furthermore, we investigate the conditions under which the classical Bhattacharyya bound becomes non-computable, deriving the necessary and sufficient criteria for the existence of these bounds. Notably, we demonstrate that the quantum Bhattacharyya bound has more relaxed conditions for computability compared to its classical counterpart.

This article is organized as follows: In Sec. \ref{Sec2}, we introduce the Bhattacharyya bounds, discuss why they are tighter bounds, and highlight scenarios in which they are advantageous. In Sec. \ref{Sec3}, we explore cases where the Bhattacharyya bound diverges and becomes non-computable, and we establish the necessary conditions for the existence of these bounds in both classical probability distributions and quantum systems. In Sec. \ref{Sec4}, we analyze the Mach-Zehnder interferometer, comparing the precision scaling achieved by the Cramér-Rao bound and the Bhattacharyya bounds. Finally, we summarize our findings and present our conclusions.

\section{Estimation theory}\label{Sec2}

To extract information from a physical system, measurements must be performed. The measurement outcome is modeled as a random variable that follows a probability distribution dependent on the state of the system. The information is encoded in this random variable, and estimation theory is the branch of statistics that addresses optimal protocols and the ultimate limits for information extraction. The Cramér-Rao bound (CRB) establishes a limit on the precision of estimating a quantity or parameter.
Consider that the outcome of the experiment, denoted by the random variable \(X\), follows the family of probability distributions \(P_{\theta}(x)\), parametrized by the parameter \(\theta\), which is being estimated. An estimator \(\Tilde{\Theta}(x)\) is a function of the random measurement outcome that provides an estimate of the parameter \(\theta\). The performance of an estimator is characterized by its bias \(b(\theta)\), mean squared error (MSE), and variance \((\Delta\Tilde{\Theta})^2\).
    \begin{align}
        b(\theta)&=\sum_x (\theta-\Tilde{\Theta}(x))P_{\theta}(x)\: ,\\
        (\Delta\tilde{\Theta})^2&=\sum_x (\langle\Tilde{\Theta}\rangle_{\theta}-\Tilde{\Theta}(x))^2P_{\theta}(x)\:,\\
        \text{MSE}&=\sum_x (\theta-\Tilde{\Theta}(x))^2P_{\theta}(x)\:.
    \end{align}
    The mean value is defined as $\langle f\rangle_\theta=\sum_x f (x)P_{\theta}(x)\: $. An estimator $\Tilde{\Theta}_{\theta_0}(x)$ is locally unbiased on the point $\theta_0$ when $\langle \Tilde{\Theta}_{\theta_0}\rangle_{\theta_0}=\theta_0$ and $\frac{d b(\theta)}{d\theta}|_{\theta=\theta_0}=0$. The subindex $\theta_0$ is used to remind that the estimator is unbiased at that point. \color{black} Note that for unbiased estimators the MSE and the variance coincide. Unbiased Cramér-Rao, usually named as the Cramér-Rao bound (CRB), states that for any unbiased estimator
     \begin{equation}
        (\Delta\tilde{\Theta}_{\theta_0})^2\geq\frac{1}{F_C},
         \label{CRB}
     \end{equation}
     where $F_C$ is the classical Fisher information and is given by
     \begin{equation}
         F_C=\sum_{x \in X_{+}}  \left(\frac{\partial P_{\theta}}{\partial\theta}\right)^2\frac{1}{P_\theta(x)}\: ,
     \end{equation}
 where $X_+$ is the space of events $x$ with nonzero probability $P_\theta(x)>0$. Previous definitions hold for discrete probability distributions but can be generalized for continuous distributions replacing sums with integrals.
     
The Cramér-Rao bound (CRB) can always be saturated in the asymptotic limit or by using a specific local estimator that depends on \(\theta\)~\cite{demkowicz2020multi}. However, in realistic scenarios where \(\theta\) is unknown, the CRB does not serve as a tight bound. Bayesian parameter estimation, in contrast, is a global approach, as Bayesian-inspired estimators do not rely on local knowledge of \(\theta\). Unfortunately, Bayesian methods are often challenging to compute~\cite{rubio2024first}.

\subsection{Bhattacharyya bound}

Bhattacharyya bounds (BhB) were introduced in the context of classical frequentist parameter estimation problems to provide a lower bound for the variance of an estimate~\cite{bha}. The BhB gives a tighter bound than the Cramér-Rao bound (CRB) by incorporating higher-order derivatives of the probability distribution in the calculation. It is shown that the BhB converges to the variance of the best unbiased estimator---the estimator that is unbiased for any point in the region of the allowed values of the parameter---when the  probability \color{black}distribution belongs to an exponential family of distributions~\cite{bli}. 

The \(n^{\text{th}}\)-order BhB gives the lowest value of the variance for an estimator that satisfies 
\[
\frac{d^i b(\theta)}{d \theta^i} \bigg|_{\theta=\theta_0} = 0, \quad \text{for } 1 \leq i \leq n,
\]
where \(b(\theta)\) denotes the bias. This estimator is unbiased over a larger region than that obtained using the CRB, as will be shown later, and its variance is closer to the actual variance in realistic parameter estimation problems---namely, cases where the true value of the parameter is unknown. The BhB is given by~\cite{Gessner},
\begin{equation}\label{eq:BhB-maximization}
    (\Delta\tilde{\Theta}_{\theta_0})^2\geq  \max _{\boldsymbol{a}} \frac{\left(\boldsymbol{a}^{\top} \boldsymbol{\lambda}\right)^2}{\boldsymbol{a}^{\top} \boldsymbol{C} \boldsymbol{a}}=\boldsymbol{\lambda}^{\top} \boldsymbol{C}^{-1} \boldsymbol{\lambda}. 
\end{equation}
The last equality is valid only when $\boldsymbol{C}^{-1}$ exists; whereas if this is not fulfilled, one needs to resort to the maximization problem. The vectors $\boldsymbol{\lambda}$ and $\boldsymbol{a}$ have $n$ entries and $\boldsymbol{C}$ is a $n\times n$ matrix, where $n$ is the order of the bound we are computing. They are defined as~\cite{Gessner}
\begin{align}
    \boldsymbol{\lambda}=\begin{pmatrix}
 1 \\
0\\
\vdots\\
0
\end{pmatrix},
\:\:\:\:\:\:\:\boldsymbol{C}_{kl}=\sum_{x \in X_{+}} \frac{\partial^kP_{\theta_0}(x) \partial^l P_{\theta_0}(x)}{P_{\theta_0}(x)},
\label{BhBmatrix}
\end{align}
where $$\partial^kP_{\theta_0}(x)=\frac{\partial^k P_{\theta}(x)}{\partial \theta^k}\bigg|_{\theta=\theta_0}.$$
The bound is in principle saturable by an estimator that satisfies,
\begin{align}
    \boldsymbol{\lambda}^\top \boldsymbol{C}^{-1} \boldsymbol{g}(x) = P_{\theta_0}(x ) \left( \tilde{\Theta}_{\theta_0}(x) - \theta_0 \right)
    \label{estimatorcl}
\end{align}
and $\boldsymbol{g}(x)^T=(\partial P_{\theta_0}(x),\partial^2P_{\theta_0}(x),...,\partial^nP_{\theta_0}(x))$. 

For a limited number of measurements, Bayesian estimation techniques are often considered more appropriate than frequentist methods (see, e.g., Refs.~\cite{e20090628,rubio2018non} for discussion). In the Bayesian framework, one assumes a prior distribution \(p_{\text{prior}}(\sigma)\) that reflects knowledge about the parameter \(\sigma\) before any measurements are made. The prior is then updated according to Bayes' rule, resulting in the Bayesian posterior distribution \(p_{\text{post}}(\sigma|x)\) after observing the measurement outcome \(x\), given by
\begin{equation}
    p_{\text{post}}(\sigma | x) = \frac{p(x | \sigma) \, p_{\text{prior}}(\sigma)}{\int \mathrm{d}\theta \, p(x | \sigma) \, p_{\text{prior}}(\sigma)}.
\end{equation}
The posterior distribution is used to induce an estimation for $\sigma$, given the outcome $x$.
\subsection{Application: Gaussian Distribution}
We consider the problem of estimating the variance $\sigma^2$ \color{black}of a zero mean Gaussian probability distribution,
\begin{align}
    P_{\sigma}(x)=\frac{1}{\sqrt{2\pi}\sigma}e^{\frac{-x^2}{2\sigma^2}}.
\end{align}

\begin{figure}
 \begin{subfigure}{0.45 \textwidth}
     \includegraphics[width=\textwidth]{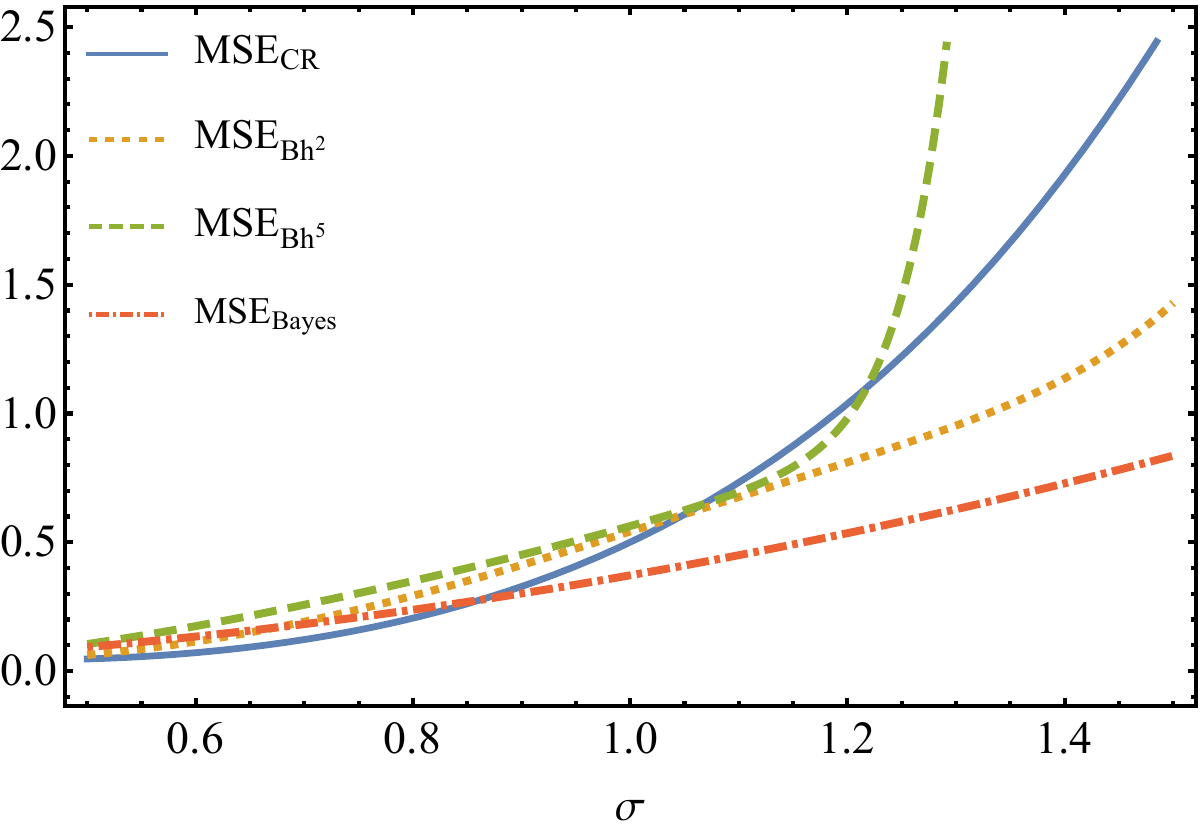}
       \caption{Mean squared error of different strategies optimized for $\sigma_0=1$ when the true value of the parameter is $\sigma$. }
     \label{fig:a}
 \end{subfigure}
 \hfill
 \begin{subfigure}{0.45\textwidth}
     \includegraphics[width=\textwidth]{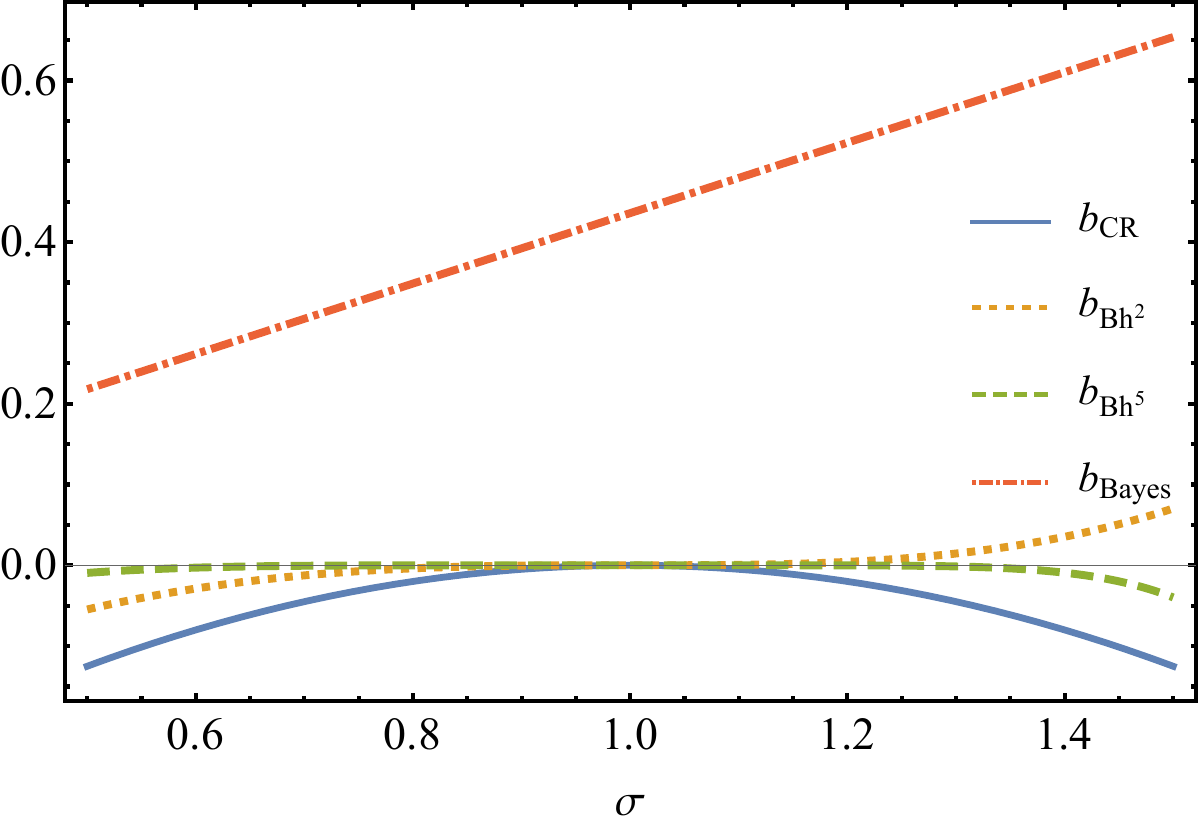}
     \caption{Bias given by different strategies optimized for $\sigma_0=1$ when the true value of the parameter is $\sigma$.}
     \label{jjj}
 \end{subfigure}
 \caption{Comparison of the MSE (upper figure) and the bias (lower figure) for a Gaussian distribution using Bayesian, CR, BhB2 and BhB5 strategies. }
    \label{fig:Bayesian}
\end{figure}

We compare the mean squared error and bias of several estimators saturating Bhattacharyya bounds for the point $\sigma_0=1$ and given by (\ref{estimatorcl}) with the estimator saturating Cramér-Rao in the same point and a Bayesian estimator. The Bayesian estimator considered
is the value of $\sigma$ that maximizes the posterior, that is, 
\begin{equation}
    \tilde{\sigma}(x)=\arg \max_{\sigma}   p_{\text{post}}(\sigma | x).
\end{equation} 

In the Bayesian approach, the information about \( \theta \) provided by the posterior probability always depends on the chosen prior distribution \( p_{\text{prior}}(\theta) \). If prior knowledge about \( \theta \) is available, it can be incorporated by selecting an appropriate prior distribution. However, in the absence of prior information, determining a "noninformative" prior becomes challenging. A flat prior, \( p_{\text{prior}}(\theta) = \text{const} \), may appear noninformative, but it is not invariant under reparameterizations, meaning it could encode unintended information for different functions of \( \theta \).

To address this, we adopt Jeffreys prior, which is given by \( p_{\text{prior}}(\theta) \propto \sqrt{F_C} \), where \( F_C \) is the classical Fisher information. Jeffreys prior is particularly useful as it remains invariant under reparameterization, ensuring consistency across different representations of the parameter.
In Fig.~\ref{fig:Bayesian}, we depict the mean squared error (MSE) and the bias of the different estimators. As expected, the Bayesian approach gives a lower MSE over a large range of values of \(\sigma\), but its bias is the largest. On the other hand, the estimator satisfying the order-5 Bhattacharyya bound (BhB) has the smallest bias in the interval considered, but its MSE increases rapidly for large values of \(\sigma\). 
\subsection{Quantum bounds}
In quantum parameter estimation~\cite{paris}, a family of quantum states \(\rho_{\theta}\), parameterized by a parameter \(\theta\), along with a particular POVM measurement \(\{\Pi_x\}\), replaces the family of probability distributions \(P_{\theta}(x)\). The lower bound on the mean squared error (MSE) of the estimator is now determined by the quantum Cramér-Rao bound (QCRB)~\cite{paris,braunstein1994statistical,mehboudi2019thermometry}. Its version for locally unbiased estimators is given by
\begin{equation}
    (\Delta\tilde{\Theta}_{\theta})^2\geq\frac{1}{F_Q},
\end{equation}
where $F_Q=\operatorname{Tr}[\rho_\theta L^2]$ is the quantum Fisher information. The symmetric logarithmic derivative (SLD) is the operator $L$ satisfying 
\begin{equation}
    \frac{d\rho_\theta}{d\theta}=\frac{\rho_\theta L+L\rho_\theta}{2}.
    \label{SLD}
\end{equation}
From Eq. \eqref{SLD} it is deduced that the SLD is a hermitian operator and its eigenvectors form an orthonormal basis. In the derivation of the QCRB, a general POVM $\{\Pi_x\}$ was considered but it was shown that a projection on the basis of eigenvectors of $L$ is the optimal measurement that saturates the QCRB~\cite{mehboudi2019thermometry}. As in the classical estimation theory, the QCRB is always saturable in the asymptotic regime. Furthermore, it has recently been shown to be non-saturable within a Bayesian perspective~\cite{gorecki2020pi,jarzyna2015true}.
 BhB can be expanded to its quantum
counterpart QBhB, as an example of bounds in quantum estimation theory~\cite{Gessner}.
\color{black}Doing so, the QBhB state that the variance of any unbiased estimator with null $n$ first bias derivatives satisfies
\begin{equation}
    (\Delta\hat{\Theta}_{\theta_0})^2\geq  \max _{\boldsymbol{a}} \frac{\left(\boldsymbol{a}^{\top} \boldsymbol{\lambda}\right)^2}{\boldsymbol{a}^{\top} \boldsymbol{Q} \boldsymbol{a}}=\boldsymbol{\lambda}^{\top} \boldsymbol{Q}^{-1} \boldsymbol{\lambda}.
    \label{10}
\end{equation}
Again, the last equality holds when $\boldsymbol{Q}^{-1}$ exists.  In Ref.~\cite{mihailescu2025metrological} the authors study the consequences of non invertible matrices in quantum multiparameter metrology. \color{black} The $n\times n$ matrix $\boldsymbol{Q}$ is defined as
\begin{align}
\boldsymbol{Q}_{kl}=\operatorname{Tr}\left (\frac{d^k\rho_k}{d\theta^k}L_l\right),  
\end{align}
where $L_l$ is the generalization of the symmetric logarithmic derivative that obeys
\begin{align}
\frac{d^l\rho_{\theta}}{d\theta^l}=\frac{\rho_{\theta}L_l+L_l\rho_{\theta}}{2}.
\end{align}
The bound in Eq. \eqref{10} is saturated by the estimator that satisfies 
\begin{align}
    \boldsymbol{\lambda}^\top \boldsymbol{Q}^{-1} \boldsymbol{g}(x) = P_{\theta_0}(x ) \left( \tilde{\Theta}_{\theta_0}(x) - \theta_0 \right),
    \label{estimator}
\end{align}
 where as in the classical case $\boldsymbol{g}(x)^T=(\partial P_{\theta_0}(x),\partial^2P_{\theta_0}(x),...,\partial^nP_{\theta_0}(x))$.
\color{black}
\section{Conditions of the existence of estimators}\label{Sec3}
Up to this point, we have presented the usefulness of the Bhattacharyya bound (BhB) for experimentally friendly scenarios. However, one cannot always construct estimators that fulfill the conditions required to lower bound their mean squared error (MSE) by the BhB. This is particularly true when dealing with discrete probability distributions. 

In this section, we provide the conditions for the existence of such estimators for finite-support probability distributions. We conclude that, in some cases, higher-order BhB may not be computable or relevant. The largest computable BhB that provides new information is related to the dimension of the support of the probability distribution. Furthermore, we show that the BhB diverges if and only if no estimator satisfies the unbiasedness conditions.

In quantum parameter estimation, given a quantum state \(\rho_\theta\), the estimation process involves two steps: performing a measurement using a POVM and applying classical post-processing to the measurement outcomes. The efficiency of the estimation protocol depends on both the choice of the POVM and the method used to process the measurement data. Consequently, optimal quantum parameter estimation involves tackling two key optimization tasks: (i) optimizing over all possible observables or, more generally, all feasible POVMs that can be implemented, and (ii) optimizing over all potential estimators \(\tilde{\Theta}\) derived from the measurement outcomes. 

While the classical Cramér-Rao bound addresses the second optimization, the first introduces a uniquely quantum aspect to the problem. In this work, we highlight the implications of this additional freedom in selecting the POVM.

\subsection{Classical}\label{3A}
We consider a family of $N$-point support probability distributions $P_{\theta}(x)$ where $P_{\theta}(x_i)\geq0$ for $x_i\in \{x_1,..,x_N\}$. The family of probability distributions is parametrized by the parameter $\theta$ that we would like to infer.\\
The most general way to write an estimator $\tilde{\Theta}(x)$ is,
\[   
\tilde{\Theta}(x) = 
     \begin{cases}
       \tilde{\theta}_1 &\quad\text{if   } x=x_1,\\
       \tilde{\theta}_2 &\quad\text{if   } x=x_2,\\
       \vdots&\quad \vdots\\
       \tilde{\theta}_N &\quad\text{if   } x=x_N.\\
      
     \end{cases}.
\]
We want to compute an unbiased estimator in the point $\theta_0$ and impose the first $n$ derivatives of the bias to be zero in the point $\theta_0$. Given that $ b(\theta)=\langle \Tilde{\Theta}\rangle_{\theta} -\theta$ the conditions are
\begin{align}
\begin{split}
    \langle\Tilde{\Theta}\rangle_{\theta_0} &=\theta_0,\\
     \frac{d \langle\Tilde{\Theta}\rangle_{\theta_0}}{d \theta}\bigg|_{\theta=\theta_0}&=1,\\
      \frac{d^l  \langle\Tilde{\Theta}\rangle_{\theta_0}}{d \theta^l}\bigg|_{\theta=\theta_0}&=0.\:\:\:\:\:\:\:\:\: l=2,3,...,n.\\
      \label{condiciones}
\end{split}  
\end{align}
Since we are considering discrete probability distributions the mean value can be written as finite sum $\langle\Tilde{\Theta}\rangle_{\theta_0} =\sum_{i}P_{\theta_0}(x_i)\Tilde{\Theta}(x_i)$. Defining $\Tilde{\Theta}(x_i)=\Tilde{\theta_i}$, Eq. \eqref{condiciones} can be written as a system of equations $A \Vec{x}=\vec{b} $ where 
\begin{align*}
A=
\begin{pmatrix}
P_{\theta_0}(x_1) &P_{\theta_0}(x_2) & \dots &  1-\sum_i^{N-1}P_{\theta_0}(x_i)  \\
\partial P_{\theta_0}(x_1) & \partial P_{\theta_0}(x_2)& \dots& -\sum_i^{N-1}\partial P_{\theta_0}(x_i)\\
\partial^2 P_{\theta_0}(x_1) & \partial^2 P_{\theta_0}(x_2)& \dots& -\sum_i^{N-1}\partial^2 P_{\theta_0}(x_i)\\
\vdots & \vdots & \ddots & \vdots \\
\partial^n P_{\theta_0}(x_1) & \partial^n P_{\theta_0}(x_2)& \dots& -\sum_i^{N-1}\partial^n P_{\theta_0}(x_i)\\\\
\end{pmatrix}
\label{system}
\end{align*}
and 
\begin{align*}
\vec{x}=&
\begin{pmatrix}
 \Tilde{\theta}_1  \\
\Tilde{\theta}_2\\
\vdots\\
\Tilde{\theta}_N
\end{pmatrix},\quad \quad \quad
\Vec{b}=\begin{pmatrix}
 \theta_0 \\
1\\
0\\
\vdots\\
0
\end{pmatrix},
\end{align*}
where we defined \( \partial^l P_{\theta_0}(x) \equiv \frac{\partial^l P_{\theta}(x)}{\partial \theta^l} \big|_{\theta=\theta_0} \). The system has \(n+1\) equations and \(N\) variables.
 Moreover, we define $A_j$ as the vector made of the elements of the $j$-th row of $A$, i.e.
\begin{align*}
A_j=
\begin{pmatrix}
\partial^j P_{\theta_0}(x_1) & \partial^j P_{\theta_0}(x_2)& \dots& -\sum_i^{N-1}\partial^j P_{\theta_0}(x_i)\\
\end{pmatrix}.
\end{align*}
We now consider the case where the set of vectors $\{A_j :1<j\leq N \}$ is linearly independent. \color{black} This implies that if \(n \leq N-1\), there is always a solution. 

If \(n > N-1\), then the higher derivatives are no longer independent and can be written as a combination of the first \(N-1\) linearly independent derivatives. In this case, the system has no solution, or the solution is trivial. That is to say, the \(n^{\text{th}}\)-order BhB is equal to the \((N-1)\)-order BhB for all \(n > N-1\). 

It is then concluded that for a probability distribution with \(N\) points, it is advisable to compute the BhB bound up to order \(N-1\). In the appendix~\ref{sec:app}, the conditions for the existence of the estimator are given. It is also shown that the bound diverges if and only if the estimator does not exist. 
\subsection{Quantum}
 We extend the previous results to determine how the maximal degree --that gives nontrivial information-- of the QBhB is limited by the dimension of quantum system. The consequences of finite-dimensional systems in parameter estimation have also been studied Ref.~\cite{gebhart2024fundamental,candeloro2024dimension}. \color{black}
In quantum estimation theory the measurement is a POVM $\{\Pi_{i}\}$, where $i$ are the possible  outcomes, each one with probability $P_{\theta}(i)=\operatorname{Tr}(\Pi_{i}\rho_\theta)$. The classical processing of the measurement outcomes is accounted by the estimator $\Tilde{\Theta}(x)$ that associates an estimate $\Tilde{\Theta}(i)$ to the outcome $i$. Every estimate $\Tilde{\Theta}(i)$ has a probability given by $P_{\theta}(i)$. From now on we are going to consider projective value measurement (PVM). These measurements are optimal in the sense that they saturate the CRB, i.e. there is always a projective measurement that satisfies $F_C=F_Q$. This measurement consists in projecting onto the eigenvectors of $L$. Analogously, it does always exist a projective measurement that saturates BhB~\cite{Gessner}. We note that since we are considering PVM's the number of outcomes coincides with the dimension of the Hilbert space\color{black}. Both tasks, measurement, and classical processing, can be treated simultaneously as a hermitian operator $\hat{\Theta}=\sum_i \Tilde{\Theta}(i)\Pi_i $~\cite{meyer2023quantum}. The conditions of Eq. \eqref{condiciones} are now written in terms of the operator $\hat{\Theta}$ as
\begin{align}
\begin{pmatrix}
 \operatorname{Tr}[\rho_{\theta_0} \hat{\Theta}  ]\\
  \operatorname{Tr}[d \rho_{\theta_0 } \hat{\Theta} ]\\
  \operatorname{Tr}[d^2 \rho_{\theta_0}\hat{\Theta} ]\\
  \vdots\\
 \operatorname{Tr}[d^n \rho_{\theta_0}\hat{\Theta} ]\\
\end{pmatrix}=
\begin{pmatrix}
\theta_0\\
 1\\
  0\\
  \vdots\\
  0\\
\end{pmatrix},   
\end{align}
where again $$d^k\rho_{\theta_0}=\frac{d^k\rho_{\theta}}{d\theta^k}\bigg|_{\theta=\theta_0}.$$ We rewrite these conditions as
\begin{align}
\begin{pmatrix}
(\rho_{\theta_0})_{ij} \hat{\Theta}_{ji} \\
  (d \rho_{\theta_0})_{ij}\hat{\Theta}_{ji} \\
  (d^2 \rho_{\theta_0})_{ij}\hat{\Theta}_{ji} \\
  \vdots\\
(d^n \rho_{\theta_0})_{ij}\hat{\Theta}_{ji}\\
\end{pmatrix}=
\begin{pmatrix}
\theta_0\\
 1\\
  0\\
  \vdots\\
  0\\
\end{pmatrix},   
\end{align}
where we are using the Einstein summation convention and $i,j=1,2,..., N$ where $N$ is the dimension of the Hilbert space of the physical system considered.
Now we have a system of equations with complex coefficients for the complex variables $\hat{\Theta}_{ij}$. Since the estimator operator must be hermitian, not all the variables $\hat{\Theta}_{ij}$ are independent. Instead of $N^2$ independent variables, there are now $\frac{N(N+1)}{2}$ independent variables. The system of equations can be vectorized and rewritten as $A\vec{x}=\vec{b}$  where,
\begin{equation}
A=
\begin{pmatrix}
(\rho_{\theta_0})_{1} &(\rho_{\theta_0})_{2} & \dots &  (\rho_{\theta_0})_{\frac{N(N+1)}{2}}  \\
(d \rho_{\theta_0}  )_{1} & (d \rho_{\theta_0})_{2}& \dots& (d \rho_{\theta_0})_{\frac{N(N+1)}{2}}\\
(d^2 \rho_\theta )_{1} & (d^2 \rho_\theta)_{2}& \dots& (d^2 \rho_\theta)_{\frac{N(N+1)}{2}}\\
\vdots & \vdots & \ddots & \vdots \\
(d^n \rho_\theta )_{1} & (d^n \rho_\theta )_{2}& \dots& (d^n \rho_\theta )_{\frac{N(N+1)}{2}}\\
\end{pmatrix}
\label{system2}
\end{equation}
and
\begin{align*}
\vec{x}=&
\begin{pmatrix}
 \hat{\Theta}_{1}  \\
\hat{\Theta}_{2}\\
\vdots\\
\hat{\Theta}_{\frac{N(N+1))}{2}}
\end{pmatrix},\quad \quad \quad
\Vec{b}=\begin{pmatrix}
 \theta_0 \\
1\\
0\\
\vdots\\
0
\end{pmatrix}.
\end{align*}

So we have a system of equations equivalent to the one for a classical probability distribution with $\frac{N(N+1)}{2}$ support points. As before if this system of equations cannot be solved the QBhB variance will go to infinity. On the other hand, if the first $\frac{N(N+1)}{2}-1$ derivatives are linearly independent, for all $n>\frac{N(N+1)}{2}-1$ one has that $n^{th}$order QBhB is equal to $\frac{N(N+1)}{2}-1$ order QBhB. The difference between the quantum and classical cases is that in the former we have the freedom to choose a measurement. This extra freedom implies that the highest relevant QBhB bound --which gives a different bound than lower orders-- is lifted in the quantum case because $N-1<\frac{N(N+1)}{2}-1$. 

Let us elaborate further on this because, at first glance, the result may seem to be at odds with the classical case. After projective measurements, one deals with \(N\)-point probability distributions. Therefore, the results derived in the classical setting must also apply, and it might seem that there is a contradiction to what we stated here. 

To resolve this apparent contradiction, we explain why the \(N\)-order quantum Bhattacharyya bound (QBhB) need not be equal to the \((N-1)\)-order QBhB. The key point is that, to compute the \(N\)- and \((N+1)\)-order QBhB, different measurements have to be performed, resulting in different probability distributions \(P_N\) and \(P_{N+1}\). This extra freedom in the choice of measurements allows quantum systems to access more non-trivial information. In other words, the maximal order for which the QBhB provides distinct information is higher than in classical scenarios. 
Thus, since \(P_N\) and \(P_{N+1}\) are different probability distributions, there is no discrepancy with what we presented in Section~\ref{3A}.

\subsection{Illustrative example}

In this subsection, we show a case in which we see the lifting in the constraints produced by the quantum nature of the problem, meaning that we see that the quantum bounds are not trivial --giving the same value as bounds from lower orders-- for $n > N-1$, being $n$ the order of the BhB employed and $N$ the number of measurement outcomes. As in this example we deal with qubits, $N = 2$.

Consider that our state $\rho(\theta)$ is obtained through the following unitary operation $U=e^{-i\theta^2 H}$ where $H=\sigma_x$ is the $x$-Pauli matrix, and we would like to estimate $\theta$. The initial state is given by
\begin{equation}
    \rho(0)= \frac{1}{2} \mathbb{I}+ \frac{(2\lambda -1)}{2} \sigma_z,
\end{equation}
where $ \lambda \in (0,1)$ and
\begin{widetext}
\begin{equation}
    \rho(\theta)=U \rho(0) U^\dag = \left(
\begin{array}{cc}
 \left(\lambda -\frac{1}{2}\right) \cos \left(2 \theta ^2\right)+\frac{1}{2} & \frac{1}{2} i (2 \lambda -1) \sin \left(2 \theta ^2\right) \\
 i (1-2 \lambda ) \sin \left(\theta ^2\right) \cos \left(\theta ^2\right) & \frac{1}{2} \left((1-2 \lambda ) \cos \left(2 \theta ^2\right)+1\right) \\
\end{array}
\right).
\end{equation}
\end{widetext}

\begin{figure}[h]
    \centering
\includegraphics[width=0.45\textwidth]{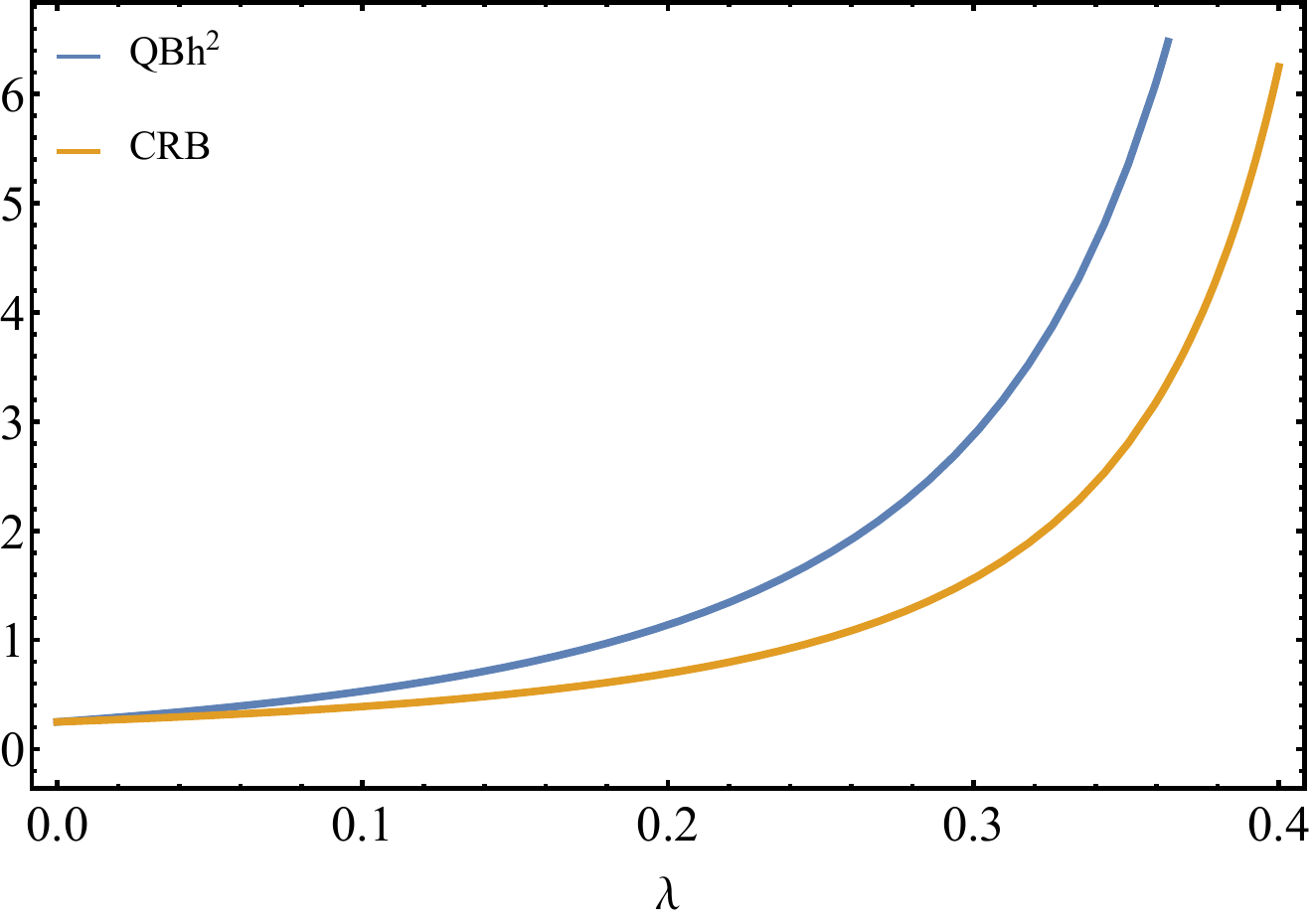}
    \caption{QCR and QBhB as a function of $\lambda$ for $\theta=1/2$. One sees that they differ, and as expected QBhB bound will be higher than QCR. In addition, we plot these values of $\lambda$ because the plot is symmetric around $\lambda=0.5$ and also divergent for this same value --since the state will be $\rho \propto \mathbb{I}$ and won't contain any information on the parameter--.}
    \label{fig:fisher-vs-BhB}
\end{figure}

We can proceed and calculate $L_1$, and $L_2$ following the definitions presented in previous sections. Doing some algebra one ends up with the QBhB matrix which reads

\begin{equation}
  \boldsymbol{Q}_{BhB^2}= \left(
\begin{array}{cc}
 16 \theta ^2 (1-2 \lambda )^2 & 16 \theta  (1-2 \lambda )^2 \\
 16 \theta  (1-2 \lambda )^2 & \frac{16 (1-2 \lambda )^2 \left((\lambda -1) \lambda -4 \theta ^4\right)}{(\lambda -1) \lambda } \\
\end{array}
\right), 
\label{B3}
\end{equation}
where the first element corresponds to the QFI. To finally get the BhB, we take the inverse of the matrix and look at the first element of it. In Fig.~\ref{fig:fisher-vs-BhB} we can see the lower bound on the variance of the estimator when using the QCR bound and QBhB. Thus, computing the \textit{quantum} Bhattacharyya bound gives new non-trivial information even though $n>N-1$.\\  

Now we consider the classical probability distributions that arise after performing a measurement. An optimal measurement to achieve the QCRB consists of a projection onto the eigenstates of \( L_1 \). On the other hand, a measurement that saturates the \( n \)-th order BhB is given by a projection onto the eigenstates of 
\[
\alpha (Q^{-1} \boldsymbol{\lambda}) \boldsymbol{L},
\]
where \( \boldsymbol{L} = (L_1, L_2, \ldots, L_n) \), \( Q \) is the BhB matrix, \( \boldsymbol{\lambda}=(1,0,..,0) \) and \( \alpha \) is a normalization constant~~\cite{Gessner}. 
The resulting probability distributions obtained by measuring the state \( \rho_\theta \) with the optimal measurement---which saturates the CRB in the point \( \phi \) are:

\begin{align*}
    P_\theta^{\text{CRB}}
(\alpha_1)&=\frac{(2 \lambda -1) \sin f-1}{\left(\sin \left(2 \phi ^2\right)-1\right) \left(\left(\tan \left(2 \phi ^2\right)+\sec \left(2 \phi ^2\right)\right)^2+1\right)}\\
P_\theta^{\text{CRB}}
(\alpha_2)&=1-P_\theta^{\text{CRB}}(\alpha_1)
\end{align*}
where $ f=2(\theta -\phi ) (\theta +\phi )$ and $\alpha_i$ are the eigenvalues of $L_1$. The probability distribution that saturates the BhB is 
\begin{widetext}
\begin{align*}
P_\theta^{Bh}
(\beta_1)=&\frac{4 \phi ^2 \big((2 \lambda -1) \left(\sin \left(2 \theta ^2-4 \phi ^2\right)-g \sin f\right)+\sin \left(2 \phi ^2\right) \left(4 (1-2 \lambda ) \phi ^2 \sin f+g\right)\big)+(1-2 \lambda ) g \cos f}{32 \phi ^4+2 g \left(4 \phi ^2 \sin \left(2 \phi ^2\right)-\cos \left(2 \phi ^2\right)\right)+2}
\\
&+\frac{\cos \left(2 \phi ^2\right) \left((2 \lambda -1) \cos f-g\right)+16 \phi ^4+1}{32 \phi ^4+2 g \left(4 \phi ^2 \sin \left(2 \phi ^2\right)-\cos \left(2 \phi ^2\right)\right)+2}
\\ 
P_\theta^{Bh}
(\beta_2)=&1-P_\theta^{Bh}
(\beta_1)
\end{align*}
\end{widetext}
where $ g= \sqrt{16 \phi ^4+1}$ and $\beta_i$ are eigenvalues of  $\alpha (Q^{-1}\boldsymbol{\lambda})\boldsymbol{L}$.
Now we compute the BhB matrix for $P_\theta^{\text{CRB}}(x)$ and for $P_\theta^{Bh}(x)$ obtaining,
\begin{equation}
  \boldsymbol{C}_{\text{CRB}}=\left(
\begin{array}{cc}
 16 \theta ^2 (1-2 \lambda )^2 & 16 \theta  (1-2 \lambda )^2 \\
 16 \theta  (1-2 \lambda )^2 & 16 (1-2 \lambda )^2 \\
\end{array}
\right)\label{B4}
\end{equation}
and
\begin{equation}
  \boldsymbol{C}_{Bh}=   \left(
\begin{array}{cc}
\frac{64 \theta ^6 (1-2 \lambda )^2}{4 \theta ^4-\lambda ^2+\lambda } & 0 \\
 0 & 0 \\
\end{array}\label{B5}
\right)\end{equation}
where we have supposed that we are doing the optimal measurement for the point $\theta$, that is, $\phi\to\theta$.
Note that $C_{\text{CRB}}$ is not invertible, using (\ref{10}) it can be noted that the bound is divergent and that there is no estimator satisfying the conditions for 2$^{nd}$order BhB. The Fisher information $F_i^{\text{CRB}}= 16 \theta ^2 (1-2 \lambda )^2$ coincide with QFI, see (\ref{B3}). 
On the other hand, although $\boldsymbol{C}_{Bh}$ is not invertible, if we resort to the maximization (\ref{10}), we obtain a finite BhB that coincides with the Cramér-Rao bound obtained for $P_\theta^{\text{Bh}}$. Thus, we have
\begin{align}
    \text{Bh}_2=\text{CRB}=\frac{4 \theta ^4-\lambda ^2+\lambda }{64 \theta ^6 (1-2 \lambda )^2}.\label{B6}
\end{align}
The QBh$_2$ derived from (\ref{B3}) coincides with the result obtained in (\ref{B6}). The key insight is that a measurement based on the eigenstates of \( (Q^{-1} \boldsymbol{\lambda}) \boldsymbol{L} \) produces a probability distribution with a null second derivative. When the second derivative vanishes, the estimator that saturates the CRB automatically satisfies the conditions for Bh$_2$. This explains why Bh$_2$ and the CRB are identical for the probability distribution \(  P_\theta^{\text{Bh}} \) and why in the quantum case one can obtain more nontrivial higher orders in the hierarchy of bounds.

\color{black}

\section{Mach-Zehnder interferometer}\label{Sec4}
Historically, greater attention has been given to the Barankin bounds~\cite{barankinthres,knockaert1997barankin}. In contrast, to the best of our knowledge, Bhattacharyya bounds have not been computed for any physical problem before. In this section, we compute the Bhattacharyya bounds and identify the estimators that achieve these bounds for the paradigmatic Mach-Zehnder interferometer. This example shows that when we move away from the point where the locally unbiased estimators are constructed, the Bhattacharyya bounds provide an improvement in the achievable precision, i.e. the MSE of the estimator that satisfies Eq.~\eqref{estimator} is smaller than the MSE of the estimator saturating Cramér-Rao.\color{black}
	\pgfdeclarelayer{background}
 \pgfdeclarelayer{foreground}
	\pgfsetlayers{background, main, foreground}
\begin{figure}
    \centering

	\begin{tikzpicture}[scale=0.75]
        \coordinate (d) at (14,7);
		\coordinate (a1) at (14,6);
		\coordinate (a1') at (14.8,6);
		\coordinate (b1) at (15,5);
		\coordinate (c1) at (15.4,5.5);
		\coordinate (a2) at (14,4);
		\coordinate (a2') at (14.80,4);
		\coordinate (b2) at (15,3);
		\coordinate (c2) at (15.4,3.5);
		\coordinate (A) at (14,5.5);
		\coordinate (O) at (14,3.5);
		\coordinate (Or) at (12,4.4);
		\coordinate (Ou) at (12,4.6);
		
		\begin{pgfonlayer}{foreground}
			\draw[laser, fill=black!20] (a1) rectangle (b1);
			\draw[laser, fill=black!40] (a1') rectangle (b1);
			\node at (c1) {$n_a$};
		\end{pgfonlayer}
		\begin{pgfonlayer}{foreground}
			\draw[laser, fill=black!20] (a2) rectangle (b2);
			\draw[laser, fill=black!40] (a2') rectangle (b2);
			\node at (c2) {$n_b$};
		\end{pgfonlayer}

		\begin{pgfonlayer}{background}
			\draw[rounded corners, thick] (10-0.4,5.5-0.4) rectangle (10+0.4,5.5+0.4);
        \node at (10,5.5) {$\varphi_a$};
		\end{pgfonlayer}
  \begin{pgfonlayer}{background}
			\draw[rounded corners, thick] (10-0.4,3.5-0.4) rectangle (10+0.4,3.5+0.4);
        \node at (10,3.5) {$\varphi_b$};
		\end{pgfonlayer}
		\begin{pgfonlayer}{foreground}
         \draw[reflector1] (12-1,4.5-0.1) rectangle (12+1,4.5+0.1);
		\end{pgfonlayer}
  \begin{pgfonlayer}{foreground}
			\draw[reflector] (8-1,4.5-0.1) rectangle (8+1,4.5+0.1);
		\end{pgfonlayer}

		\begin{pgfonlayer}{main}
			\draw[ray] (Ou) -- (A);
			\draw[ray] (Or) -- (O);
            \draw[ray] (10.4,5.5) -- (Ou);
			\draw[ray] (10.4,3.5) -- (Or);
            \draw[ray] (8,4.6) -- (10-0.4,5.5);
			\draw[ray] (8,4.4) -- (10-0.4,3.5);
            \draw[ray] (6,5.5) -- (8,4.6);
			\draw[ray] (6,3.5) -- (8,4.4);
		\end{pgfonlayer}
	
		\node at (6.6, 5.5) {$a$};
        \node at (6.6, 3.5) {$b$};
        \node at (13, 5.5) {$a^\prime$};
        \node at (13, 3.5) {$b^\prime$};
        \node at (5.5, 5.6) {\Large$\ket{r}$};
        \node at (5.5, 3.4) {\Large$\ket{r}$};
	\end{tikzpicture}
  \caption{Mach-Zehnder interferometer setup.
  A 50:50 beam splitter is illuminated by two spatially separate input modes, represented by the annihilation operators \( \hat{a} \) and \( \hat{b} \), respectively. Both input modes are initialized in identical Fock states \( |r\rangle \), with \( r \) denoting the fixed photon number in each mode. Upon interaction with the first beam splitter, the input modes are coherently mixed, producing two output modes. These output modes then propagate through paths experiencing distinct phase shifts, denoted by \( \varphi_a \) and \( \varphi_b \). The phase-modulated modes are subsequently recombined at a second 50:50 beam splitter. Finally, photon number measurements are performed on the output modes of the second beam splitter, enabling interference-based observations of the system.}
    \label{Mach}
\end{figure}
\subsubsection{Mach-Zehnder}
In the Mach-Zehnder configuration, depicted in Fig.~\ref{Mach}, two modes of the electromagnetic field interfere on a balanced beamsplitter, the output beams acquire a relative phase \(\theta = \varphi_a - \varphi_b\), and interfere again on the second beamsplitter. Finally, the photon numbers \(n_a\) and \(n_b\) are measured at the output ports. 

\begin{figure}[ht]
    \centering
    \begin{subfigure}[b]{0.45\textwidth}
        \centering
        \includegraphics[width=\textwidth]{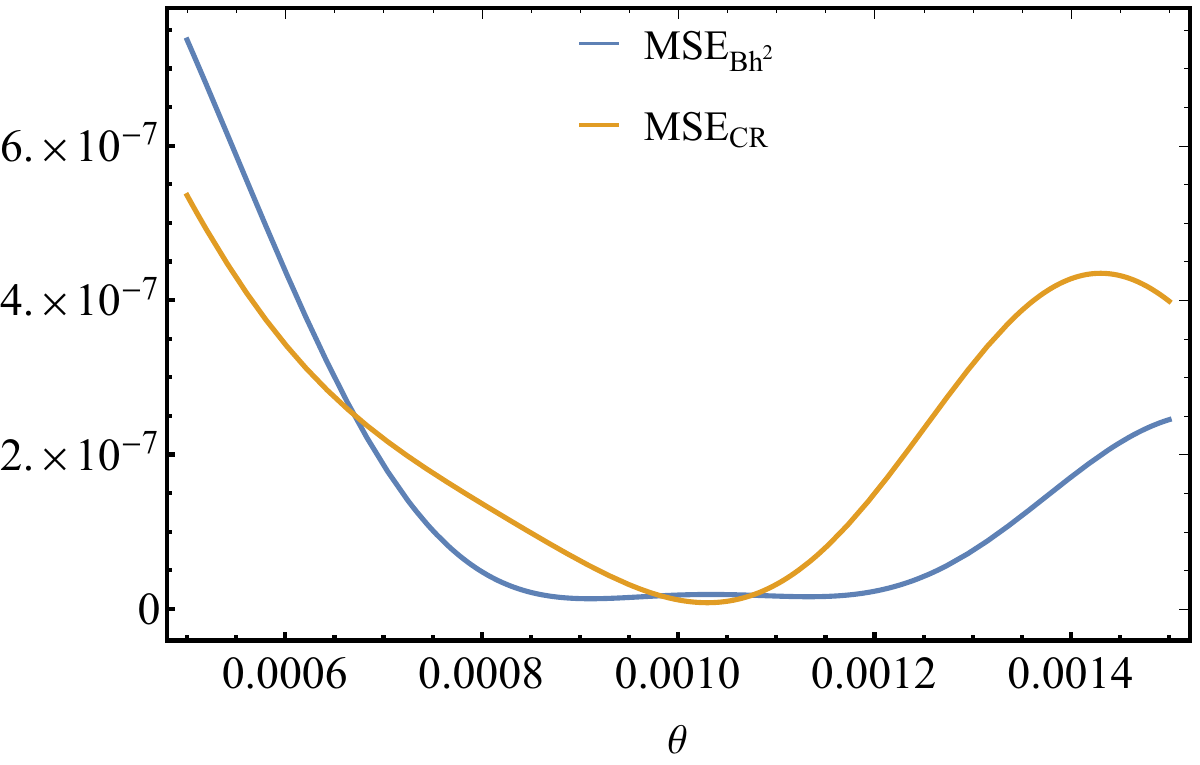}
        \label{max-acc-fisher-low}
    \end{subfigure}
    \hfill
    \begin{subfigure}[b]{0.45\textwidth}
        \centering
        \includegraphics[width=\textwidth]{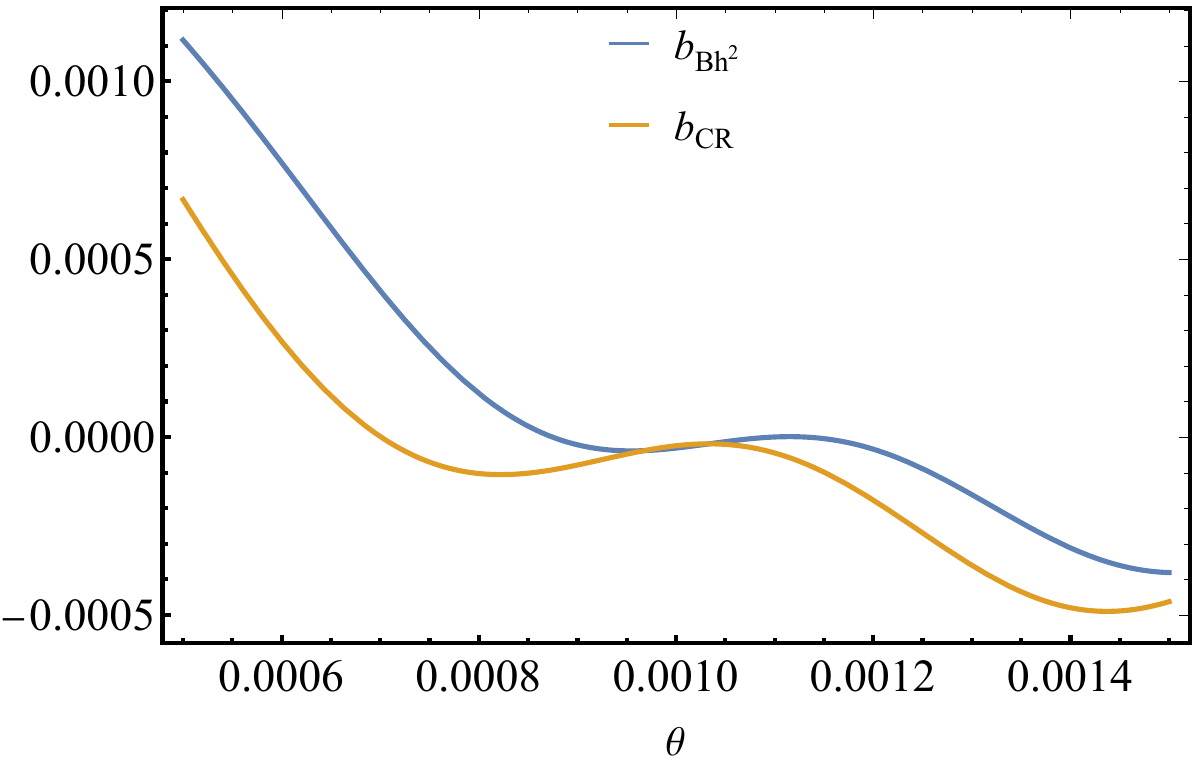}
        \label{max-acc-fisher-big}
    \end{subfigure}
    \caption{Results for the input state $\ket{r,r}$ for $r=5000$ and $\theta_0=0.001$.}
    \label{PlotMSE}
\end{figure}

We consider the input state \(\ket{r,r}\), where both input modes are in the Fock state with \(r\) photons. It has been shown that this state achieves the Heisenberg limit~\cite{holland1993} while estimating the relative phase \(\theta\). The evolution induced by the interferometer is given by the unitary operator \(U = e^{\frac{\theta}{2}(a^{\dagger}b - b^{\dagger}a)}\)~\cite{demkowicz2015quantum}. The probability distribution \(P_\theta(2q)\) is then
\begin{equation}
    P_\theta(2q)=|\langle r-q,r+q|e^{\frac{\theta}{2}(a^{\dagger}b-b^{\dagger}a)}|r,r\rangle|^2
\end{equation}
where  $2q=n_a-n_b$ is the photon counting difference. The width of this probability distribution is $|q|< r\theta$. If $\theta\ll1$ the probability distribution can be well approximated by~\cite{holland1993}
\begin{equation}
    P_\theta(2q)=J^2_q( r\theta),
\end{equation}
where $J_q(x)$ are the Bessel functions.
We compute the CRB, the \(2^{\text{nd}}\)-order BhB, and the estimators achieving each bound at \(\theta_0 = 10^{-3}\). In Fig.~\ref{PlotMSE}, we compare the mean squared error (MSE) and bias of these two estimators within an interval of \(\theta\) that contains \(\theta_0\). 

One can conclude that the BhB offers a more robust bound; i.e., \(\text{MSE}_{\text{Bh}}\) remains flat over an interval of \(\theta\), whereas \(\text{MSE}_{\text{CR}}\) increases rapidly when \(\theta \neq \theta_0\).
 Surprisingly, the behaviour of the bias of both estimators is quite similar. In terms of the bias, there is no clear advantage of using the CRB or Bh2 estimator. \color{black} 
Finally, the protocol that saturates the BhB achieves a smaller MSE. Therefore, it may be advisable to use this bound as a figure of merit instead of the CRB in certain specific scenarios.

\begin{figure}[ht]
    \centering
    \includegraphics[width=0.45\textwidth]{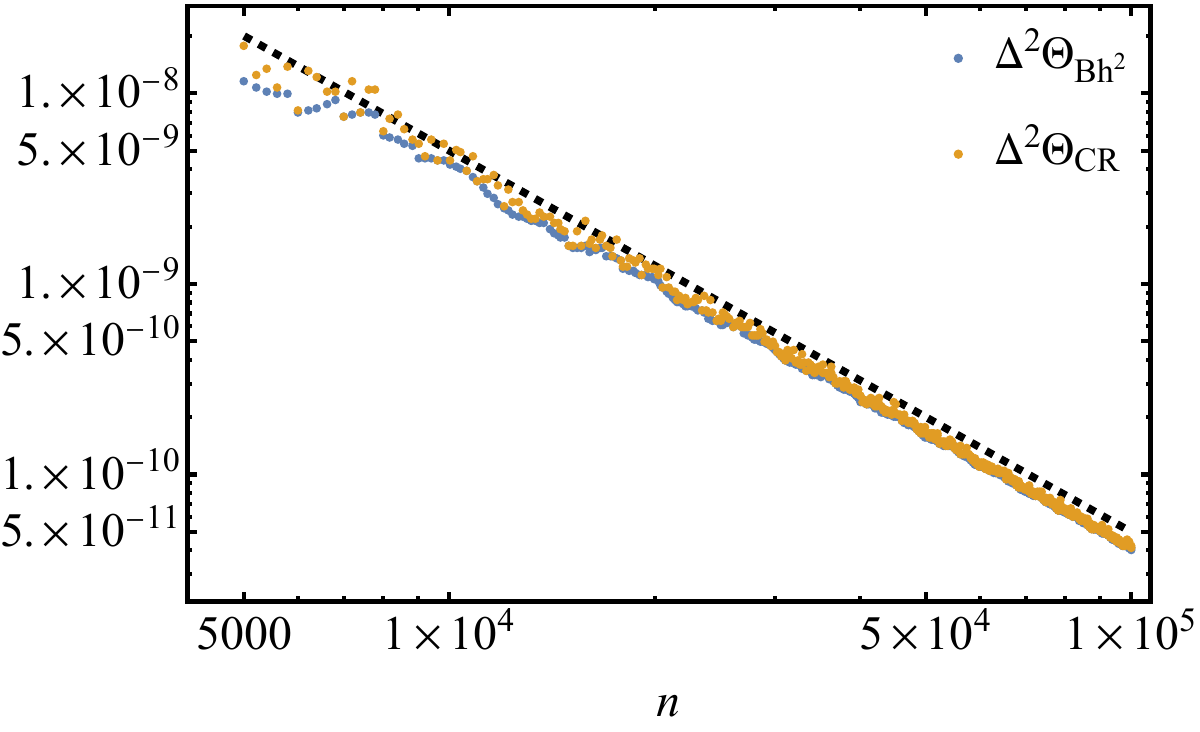}
    \caption{A logarithmic plot of the Cramér-Rao bound and the second-order Bhattacharyya bound is shown. As a reference the black dotted line is proportional to $\frac{1}{n^2}$. Both bounds have the same scaling as the dotted line, which means that both bounds attain the Heisenberg limit.\color{black}}
    \label{scale}
\end{figure}

In Fig.~\ref{scale}, we compute the second-order Bhattacharyya bound and the Cramér-Rao bound as a function of the increasing number of photons. We observe that the Cramér-Rao bound achieves the Heisenberg limit, as expected. Interestingly, the Bhattacharyya bound also reaches the Heisenberg limit, indicating that a tighter bound can still achieve Heisenberg scaling. This result suggests an alternative approach to constructing estimators that can achieve optimal scaling even  when the knowledge of the true value of the parameter is sparse. \color{black}In particular, this demonstrates that it is possible to design estimation protocols that are more robust to uncertainties in prior knowledge while still reaching the Heisenberg limit in precision.

\section{Conclusions}\label{Sec5}
Since long ago, it has been clear in estimation theory that while the Cramér-Rao bound (CRB) is extremely useful and illustrative, it is not always meaningful. For that reason, different bounds were constructed and introduced in the classical regime. On the other hand, until very recently, there were no works extending these bounds to the quantum regime~\cite{tsuda2007bhattacharyya,Gessner}.

In this article, we first motivated the use of Bhattacharyya bounds by studying its properties in several examples. Essentially, when one does not know the prior value of the parameter exactly, but knows that it is within a small range, the BhB are more desirable. We then gave necessary and sufficient conditions for an estimator that satisfies the BhB to exist. The existence conditions are related to the support of the probability distribution, which determines the freedom we have to construct an estimator. 

When we computed the \(n\)-th order Bhattacharyya bound, we found a limit for the variance of an estimator that satisfies \(n+1\) unbiasedness conditions. When the support of the probability distribution is small, we might not have enough freedom to construct an estimator satisfying the desired constraints. Namely, the estimator exists if and only if the maximization presented in Eq.~\eqref{eq:BhB-maximization} can be calculated (non-divergent). 

Moreover, following a similar reasoning, we could anticipate when computing higher orders in the BhB gives no extra information. Specifically, we saw that for \(N\) outcomes, it is ``sufficient" to calculate the \(n\)-th order BhB bound where \(n \leq N-1\). However, in the quantum case, where we have more freedom in the measurements, the order is lifted to \(n \leq \frac{N(N+1)}{2} - 1\), implying that in the quantum regime,  one can construct tighter bounds, i.e. higher order BhB. \color{black}

Finally, we compute the Bhattacharyya bound for a Mach-Zehnder interferometer and numerically show that this bound scales inversely proportional to the square of the number of photons, which is the Heisenberg scaling. Since Bhattacharyya bounds are tighter, this result supports the claim that it is possible to achieve Heisenberg scaling in the Mach-Zehnder interferometer using a more robust estimator.

\begin{acknowledgments}
We acknowledge fruitful discussions with Manuel Gessner and Maximilian Reichert. The authors acknowledge financial support from OpenSuperQ+100 (Grant No. 101113946) of the EU Flagship on Quantum Technologies, as well as from the EU FET-Open project EPIQUS (Grant No. 899368), also from Project Grant No. PID2021-125823NA-I00 595 and Spanish Ramón y Cajal Grant No. RYC-2020-030503-I funded by MCIN/AEI/10.13039/501100011033 and by “ERDF A way of making Europe” and “ERDF Invest in your Future,” this project has also received support from the Spanish Ministry for Digital Transformation and of Civil Service of the Spanish Government through the QUANTUM ENIA project call - Quantum Spain, EU through the Recovery, Transformation and Resilience Plan – NextGenerationEU within the framework of the Digital Spain 2026, and by the EU through the Recovery, Transformation and Resilience Plan-- NextGenerationEU within the framework of the Digital Spain 2026 Agenda, we acknowledge funding from Basque Government through Grant No. IT1470-22 and the IKUR Strategy under the collaboration agreement between Ikerbasque Foundation and BCAM on behalf of the Department of Education of the Basque Government. 
\end{acknowledgments}
\bibliographystyle{bibstyle}
\bibliography{apssamp}
\newpage
 \onecolumngrid
\appendix
\begin{section}{Existence of the estimator}\label{sec:app}
We study the solution of the system $A\vec{x}=\vec{b}$ with 
\begin{equation}
A=
\begin{pmatrix}
P_{\theta_0}(x_1) &P_{\theta_0}(x_2) & \dots &  1-\sum_i^{N-1}P_{\theta_0}(x_i)  \\
\partial P_{\theta_0}(x_1) & \partial P_{\theta_0}(x_2)& \dots& -\sum_i^{N-1}\partial P_{\theta_0}(x_i)\\
\partial^2P_{\theta_0}(x_1)& \partial^2P_{\theta_0}(x_2)& \dots& -\sum_i^{N-1}\partial^2P_{\theta_0}(x_i)\\
\vdots & \vdots & \ddots & \vdots \\
\partial^nP_{\theta_0}(x_1) & \partial^nP_{\theta_0}(x_2)& \dots& -\sum_i^{N-1}\partial^nP_{\theta_0}(x_i)\\
\end{pmatrix},
\quad\quad
\vec{x}=
\begin{pmatrix}
 \Tilde{\theta}_1  \\
\Tilde{\theta}_2\\
\vdots\\
\Tilde{\theta}_N
\end{pmatrix},
\quad
\quad
\Vec{b}=\begin{pmatrix}
 \theta_0 \\
1\\
0\\
\vdots\\
0
\end{pmatrix}.
\label{system1}
\end{equation}
We point out that the matrix has dimensions $(n+1)\times N $ so if $n+1>N$ there are more equations than variables and a solution might not exist. Also, it is important to stress that the derivatives inside the matrix are all evaluated at point $\theta_0$ but we do not write it for the sake of clarity.

Rouché-Frobenius theorem states that the system has a solution if and only if $\operatorname{rank}(A|\vec{b})=\operatorname{rank}(A)$. If all the rows of $A$ are linearly independent matrix $(A|\vec{b})$ wouldn't have a larger rank than $A$. Then, we are interested in the case where $A$ has linearly dependent rows. Notice that the first row is linearly independent of the others so the system $A \Vec{x}=\vec{b} $ has no solution if and only if $A^\prime \Vec{x}=\vec{\lambda} $ has no solution. From now on we study the latter system 
\begin{equation}
A^\prime=
\begin{pmatrix}
\partial P_{\theta_0}(x_1) & \partial P_{\theta_0}(x_2)& \dots& -\sum_i^{N-1}\partial P_{\theta_0}(x_i)\\
\partial^2P_{\theta_0}(x_1)& \partial^2P_{\theta_0}(x_2)& \dots& -\sum_i^{N-1}\partial^2P_{\theta_0}(x_i)\\
\vdots & \vdots & \ddots & \vdots \\
\partial^nP_{\theta_0}(x_1) & \partial^nP_{\theta_0}(x_2)& \dots& -\sum_i^{N-1}\partial^nP_{\theta_0}(x_i)\\
\end{pmatrix},
\quad \quad \Vec{x}=
\begin{pmatrix}
 \Tilde{\theta}_1  \\
\Tilde{\theta}_2\\
\vdots\\
\Tilde{\theta}_N
\end{pmatrix},
\quad \quad
\Vec{\lambda}=\begin{pmatrix}
1\\
0\\
\vdots\\
0
\end{pmatrix}.
\end{equation}

The matrix $A^{\prime}$ has dimensions $n \times N$ and since the last column is linearly dependent (is the sum of the $N-1$ first columns) there are at most $N-1$ linearly independent rows. As we mentioned before linearly dependent rows are a necessary condition for the system not having a solution. For that reason, we suppose that there is a row $m$ that can be written as,
\begin{equation}
\begin{pmatrix}
\partial^mP_{\theta_0}(x_1) & \partial^mP_{\theta_0}(x_2) & \dots& -\sum_i^{N-1}\partial^mP_{\theta_0}(x_i) \\
\end{pmatrix}=\sum_{l}\alpha_l
\begin{pmatrix}
\partial^lP_{\theta_0}(x_1) & \partial^lP_{\theta_0}(x_2) & \dots& -\sum_i^{N-1}\partial^lP_{\theta_0}(x_i) \\
\end{pmatrix}.
\label{18}
\end{equation}
Without losing generality we can consider that $m\neq 1$. Now we consider the $m$ row of the augmented matrix $(A^{\prime}|\vec{\lambda})$,
\begin{equation}
\begin{pmatrix}
\partial^mP_{\theta_0}(x_1) & \partial^mP_{\theta_0}(x_2) & \dots& -\sum_i^{N-1}\partial^mP_{\theta_0}(x_i) & 0\\
\end{pmatrix}.
\label{5}
\end{equation}
If the $m$ row of the augmented matrix Eq. \eqref{5} is linearly independent from the rows of $(A^{\prime}|\vec{\lambda})$ then $\operatorname{rank}(A^{\prime}|\vec{\lambda})>\operatorname{rank}(A^{\prime})$ and there is no solution.

Thus we investigate whether it is linearly dependent or not. According to our assumption, we can write Eq. \eqref{5} as,
\begin{equation}
\begin{pmatrix}
\partial^mP_{\theta_0}(x_1) &  \dots& -\sum_i^{N-1}\partial^mP_{\theta_0}(x_i) & 0\\
\end{pmatrix}=\sum_{l}\alpha_l
\begin{pmatrix}
\partial^lP_{\theta_0}(x_1) &  \dots& -\sum_i^{N-1}\partial^lP_{\theta_0}(x_i) & 0\\
\end{pmatrix}
\end{equation}
but the first element of the sum,
\begin{equation}
\alpha_1
\begin{pmatrix}
\partial P_{\theta_0}(x_1) &  \dots& -\sum_i^{N-1}\partial P_{\theta_0}(x_i) & 0\\
\end{pmatrix}
\end{equation}
is not a row of the matrix $(A^{\prime}|\vec{\lambda})$ so Eq. \eqref{5} is linearly independent when $\alpha_1\neq0$. Then, we conclude that the system given in Eq. (\ref{system1}) has no solution if and only if,
\begin{equation}
\begin{pmatrix}
\partial^mP_{\theta_0}(x_1) & \partial^mP_{\theta_0}(x_2) & \dots& -\sum_i^{N-1}\partial^mP_{\theta_0}(x_i) \\
\end{pmatrix}=\sum_{l}\alpha_l
\begin{pmatrix}
\partial^lP_{\theta_0}(x_1) & \partial^lP_{\theta_0}(x_2) & \dots& -\sum_i^{N-1}\partial^lP_{\theta_0}(x_i) \\
\end{pmatrix},
\label{8}
\end{equation}
where $\alpha_1\neq 0$.

Now we look at the Bhattacharyya bound 
\begin{equation}\label{BhB-bounds}
    (\Delta\hat{\Theta}_{\theta_0})^2\geq  \max _{\boldsymbol{a}} \frac{\left(\boldsymbol{a}^{\top} \boldsymbol{\lambda}\right)^2}{\boldsymbol{a}^{\top} \boldsymbol{C} \boldsymbol{a}}.
\end{equation}
We study the case when the BhB variance is unbounded. It is unbounded if and only if there is a vector $\boldsymbol{a}^{\prime}$ that satisfies $\boldsymbol{C} \cdot\boldsymbol{a}^{\prime}=0$ and $\boldsymbol{a}^{\prime\top} \boldsymbol{\lambda}\neq0$. This vector $\boldsymbol{a}^{\prime}$ exists if and only if the next system of equations
\begin{equation}
\begin{pmatrix}
\sum_i\frac{1}{P_{\theta_0}(x_i)}\partial P_{\theta_0}(x_i)\partial P_{\theta_0}(x_i) & \sum_i\frac{1}{P_{\theta_0}(x_i)}\partial P_{\theta_0}(x_i) \partial^2 P_{\theta_0}(x_i)  &\dots&\sum_i\frac{1}{P_{\theta_0}(x_i)}\partial P_{\theta_0}(x_i) \partial^n P_{\theta_0}(x_i)   \\
\sum_i\frac{1}{P_{\theta_0}(x_i)}\partial^2 P_{\theta_0}(x_i)\partial P_{\theta_0}(x_i) & \sum_i\frac{1}{P_{\theta_0}(x_i)}\partial^2 P_{\theta_0}(x_i) \partial^2 P_{\theta_0}(x_i)  &\dots&\sum_i\frac{1}{P_{\theta_0}(x_i)}\partial^2 P_{\theta_0}(x_i) \partial^n P_{\theta_0}(x_i)   \\
\vdots & \vdots&\ddots&\vdots   \\
\sum_i\frac{1}{P_{\theta_0}(x_i)}\partial P_{\theta_0}(x_i)\partial^n P_{\theta_0}(x_i) & \sum_i\frac{1}{P_{\theta_0}(x_i)}\partial^n P_{\theta_0}(x_i) \partial^2 P_{\theta_0}(x_i)  &\dots&\sum_i\frac{1}{P_{\theta_0}(x_i)}\partial^n P_{\theta_0}(x_i) \partial^n P_{\theta_0}(x_i)   \\
\end{pmatrix}\begin{pmatrix}
k\\
a_2\\
\vdots\\
a_{n}
\end{pmatrix}=0,
\label{CBhB}
\end{equation}
has a solution and also $k\neq0$ because the condition that $\boldsymbol{a}^{\prime\top} \boldsymbol{\lambda}\neq0$  has to be fulfilled too.

The system of equations can be rewritten as, 
\begin{align*}
\begin{pmatrix}
 \sum_i\frac{1}{P_{\theta_0}(x_i)}\partial P_{\theta_0}(x_i) \partial^2 P_{\theta_0}(x_i)  &\dots&\sum_i\frac{1}{P_{\theta_0}(x_i)}\partial P_{\theta_0}(x_i) \partial^n P_{\theta_0}(x_i)   \\\sum_i\frac{1}{P_{\theta_0}(x_i)}\partial^2 P_{\theta_0}(x_i) \partial^2 P_{\theta_0}(x_i)  &\dots&\sum_i\frac{1}{P_{\theta_0}(x_i)}\partial^2 P_{\theta_0}(x_i) \partial^n P_{\theta_0}(x_i)   \\
 \vdots&\ddots&\vdots   \\
 \sum_i\frac{1}{P_{\theta_0}(x_i)}\partial^n P_{\theta_0}(x_i) \partial^2 P_{\theta_0}(x_i)  &\dots&\sum_i\frac{1}{P_{\theta_0}(x_i)}\partial^n P_{\theta_0}(x_i) \partial^n P_{\theta_0}(x_i)   \\
\end{pmatrix}\begin{pmatrix}
a_2\\
\vdots\\
a_{n}
\end{pmatrix}= k 
\begin{pmatrix}
\sum_i\frac{1}{P_{\theta_0}(x_i)}\partial P_{\theta_0}(x_i)\partial P_{\theta_0}(x_i)\\
\sum_i\frac{1}{P_{\theta_0}(x_i)}\partial P_{\theta_0}(x_i)\partial^2 P_{\theta_0}(x_i)\\
\vdots\\
\sum_i\frac{1}{P_{\theta_0}(x_i)}\partial P_{\theta_0}(x_i)\partial^n P_{\theta_0}(x_i)
\end{pmatrix}.
\end{align*}
Rouché-Frobenius theorem states that the system has a solution if and only if $\operatorname{rank}(A|\vec{b})=\operatorname{rank}(A)$. This is equivalent to say that $\vec{b}$ is a linear combination of columns of the matrix $(A)$, i.e. \\
\begin{align*}
\begin{pmatrix}
\sum_i\frac{1}{P_{\theta_0}(x_i)}\partial P_{\theta_0}(x_i)\partial P_{\theta_0}(x_i)\\
\vdots\\
\sum_i\frac{1}{P_{\theta_0}(x_i)}\partial P_{\theta_0}(x_i)\partial^n P_{\theta_0}(x_i)
\end{pmatrix}=\sum_{l\neq1} \alpha_l
\begin{pmatrix}
 \sum_i\frac{1}{P_{\theta_0}(x_i)}\partial^l P_{\theta_0}(x_i)\partial P_{\theta_0}(x_i)\\
\vdots\\
\sum_i\frac{1}{P_{\theta_0}(x_i)}\partial^l P_{\theta_0}(x_i)\partial^n P_{\theta_0}(x_i)
\end{pmatrix}.
\end{align*}
Rearranging the previous expression,
\begin{align}
\begin{pmatrix}
\sum_i\frac{1}{P_{\theta_0}(x_i)}\partial^m P_{\theta_0}(x_i)\partial P_{\theta_0}(x_i)\\
\vdots\\
\sum_i\frac{1}{P_{\theta_0}(x_i)}\partial^m P_{\theta_0}(x_i)\partial^n P_{\theta_0}(x_i)
\end{pmatrix}=\sum_{l=1,l\neq m} \alpha^{\prime}_l
\begin{pmatrix}
\sum_i\frac{1}{P_{\theta_0}(x_i)}\partial^l P_{\theta_0}(x_i)\partial P_{\theta_0}(x_i)\\
\vdots\\
\sum_i\frac{1}{P_{\theta_0}(x_i)}\partial^l P_{\theta_0}(x_i)\partial^n P_{\theta_0}(x_i)
\end{pmatrix}.
\label{12}
\end{align}
where $\alpha_1^{\prime}\neq0$.
That is, there is a column $m$ that is a combination of the first column and other columns of $\boldsymbol{C} $ .
Note that the elements of the $l$ column of the rhs of Eq. \eqref{12} can be written as,
\begin{align*}
\begin{pmatrix}
\partial^l P_{\theta_0}(x_1)&\partial^l P_{\theta_0}(x_2)&\dots&\partial^l P_{\theta_0}(x_N)\\
\end{pmatrix}.\begin{pmatrix}
 \frac{1}{P_{\theta_0}(x_1)}\partial^j P_{\theta_0}(x_1)\\
\frac{1}{P_{\theta_0}(x_2)}\partial^j P_{\theta_0}(x_2)\\
\vdots\\
\frac{1}{P_{\theta_0}(x_N)}\partial^j P_{\theta_0}(x_N)\\
\end{pmatrix},
\end{align*}
where $j$ goes from 1 to $n$ for each element of the column. Defining
\begin{align}
A_m&=
\begin{pmatrix}
\partial^m P_{\theta_0}(x_1)&\partial^m P_{\theta_0}(x_2)&\dots&\partial^m P_{\theta_0}(x_N)\\
\end{pmatrix},\label{B11}\\
B_j&=\begin{pmatrix}
\frac{1}{P_{\theta_0}(x_1)}\partial^j P_{\theta_0}(x_1)\\
\frac{1}{P_{\theta_0}(x_2)}\partial^j P_{\theta_0}(x_2)\\
\vdots\\
\frac{1}{P_{\theta_0}(x_N)}\partial^j P_{\theta_0}(x_N)\\
\end{pmatrix}.
\end{align}

Equation \eqref{12} can be written as,
\begin{align}
    A_m\cdot B_j&=\sum_{l=1}\alpha_l\: A_l\cdot B_j\quad \forall j .
    \label{B12}
\end{align}
This implies
\begin{align}
    A_m&=\sum_{l=1}\alpha_l\: A_l.
    \label{BB14}
\end{align}
Substituting the definitions given in  Eq. \eqref{B11} in Eq. \eqref{BB14} we get
\begin{equation}
\begin{pmatrix}
\partial^mP_{\theta_0}(x_1) & \partial^mP_{\theta_0}(x_2) & \dots& -\sum_i^{N-1}\partial^mP_{\theta_0}(x_i) \\
\end{pmatrix}=\sum_{l}\alpha_l
\begin{pmatrix}
\partial^lP_{\theta_0}(x_1) & \partial^lP_{\theta_0}(x_2) & \dots& -\sum_i^{N-1}\partial^lP_{\theta_0}(x_i) \\
\end{pmatrix},
\label{13}
\end{equation}

We have proved that BhB is unbounded if and only if Eq. \eqref{13} is true. We note that Eq. \eqref{13} is the same as Eq. \eqref{8}.
This is the same as saying that there is no estimator satisfying the conditions. As stated before, the matrix of Eq. \eqref{system} has at most $N$ linearly independent rows associated with $N$ independent conditions. Consider the case where the first $N-1$  higher order derivatives are linearly independent, then $n^{th}$order BhB bound will be equal to the $(N-1)^{th}$ BhB bound if $n\geq N-1$. 

\end{section}



\end{document}